\documentclass[twocolumn,secnumarabic,amssymb,nobibnotes,aps,superscriptaddress,pra]{revtex4-1}
 
\usepackage[dvips]{graphicx}
\usepackage{bm}
\usepackage{latexsym} 
\usepackage{amsmath}
\usepackage{mathtools}
\usepackage{empheq}
\usepackage{cases} 
\usepackage{ulem}
\usepackage{gensymb}
\usepackage{cancel}
\usepackage[T1]{fontenc}

\usepackage{amssymb}
\usepackage{color}

\newcommand{\bmH}{{\bm H}}
\newcommand{\bmh}{{\bm h}}

\newcommand{\bmm}{{\bm m}}
\newcommand{\bmn}{{\bm n}}

\newcommand{\bmS}{{\bm S}}
\newcommand{\bmsig}{{\bm \sigma}}

\newcommand{\bra}{\langle}
\newcommand{\ket}{\rangle}
\newcommand{\kB}{k_{\rm B}}

\makeatletter
\renewcommand*{\p@subsection}{}
\renewcommand*{\p@subsubsection}{}
\makeatother

\begin{document}

\title{
  Ginzburg-Landau theory of spin pumping through an antiferromagnetic layer\\
  near the N\'{e}el temperature 
}

\author{Yuto Furutani} 
\affiliation{Department of Physics, Okayama University, Okayama 700-8530, Japan}

\author{Hayato Fukushima} 
\affiliation{Department of Physics, Okayama University, Okayama 700-8530, Japan}

\author{Yutaka Yamamoto} 
\affiliation{Department of Physics, Okayama University, Okayama 700-8530, Japan}
\author{Masanori Ichioka}
\affiliation{Research Institute for Interdisciplinary Science, Okayama University, Okayama 700-8530, Japan}
\affiliation{Department of Physics, Okayama University, Okayama 700-8530, Japan}
\author{Hiroto Adachi}
\affiliation{Research Institute for Interdisciplinary Science, Okayama University, Okayama 700-8530, Japan}
\affiliation{Department of Physics, Okayama University, Okayama 700-8530, Japan}
\date{\today}

\begin{abstract}
  Spin pumping is a microwave-driven means for injecting spins from a ferromagnet into the adjacent target material. The insertion of a thin antiferromagnetic layer between the ferromagnet and the target material is known to enhance the spin pumping signal. Here, in view of describing dynamic fluctuations of the N\'{e}el order parameter, we develop Ginzburg-Landau theory of the spin pumping in a ferromagnet/antiferromagnet/heavy metal trilayer in the vicinity of the antiferromagnetic N\'{e}el temperature $T_{\rm N}$. When there exists an interfacial exchange interaction between the ferromagnetic spins and the antiferromagnetic N\'{e}el order parameter at the ferromagnet/antiferromagnet interface, we find a strongly frequency-dependent enhancement of the pumped spin current that is peaked at $T_{\rm N}$. The present finding offers an explanation for the enhanced spin pumping with strong frequency dependence observed in a Y$_3$Fe$_5$O$_{12}$/CoO/Pt system. 
\end{abstract}



\maketitle

\section{Introduction \label{Sec:I}}
The emerging field of antiferromagnetic spintronics~\cite{Jungwirth16,Zelezny18,Baltz18,Jungfleisch18} offers a new platform to examine the interplay of N\'{e}el (staggered magnetization) order parameter and spins. Historically, this interplay has long been known since the discovery of the exchange bias effect~\cite{Meiklejohn56}, where a ferromagnetic hysteresis loop of a ferromagnet/antiferromagnet bilayer is shifted unidirectionally. While this well-established phenomenon belongs to a {\it static} effect, one of the forefronts of antiferromagnetic spintronics is the {\it dynamic} manipulation of the N\'{e}el order parameter and magnetization, such as an electrical switching of the N\'{e}el order parameter~\cite{Wadley16,Bodnar18}. Because the reversal of the N\'{e}el vector relies on spin torques~\cite{Gomonay10,Cheng14,Moriyama18} and is thus driven by a flow of spins termed spin current, a great deal of investigations on antiferromagnetic spin transport have been underway, both experimentally~\cite{Lebrun18,Yuan18,QLi19,Dabrowski20,Hortensius21,Wang22,deWal23,Wang23} and theoretically~\cite{Takei15,Khymyn16,Rezende16,Bender17,Tatara19,Shen19,Yamamoto19,Troncoso20,Aoyama20,Yamamoto22,Ishikawa23}.

Investigations of antiferromagnetic spin transport require spin injection into antiferromagnets. Along with other techniques such as spin Seebeck effect~\cite{Uchida08,Uchida10a,Xiao10,Adachi11,Adachi13,Adachi25} and spin Hall effect~\cite{Dyakonov71,Hirsch99,Kato04,Wunderlich05,Valenzuela06,Kimura07}, the spin pumping~\cite{Tserkovnyak02,Mizukami02,Saitoh06} driven by the ferromagnetic resonance is known as one of the most efficient spin injection method. In 2014, Hahn {\it et al.}~\cite{Hahn14} performed a spin pumping experiment through an antiferromagnetic NiO layer in a Y$_3$Fe$_5$O$_{12}$/NiO/Pt trilayer, where the spin current signal is detected electrically in the Pt layer by the inverse spin Hall effect. In the same year, Wang {\it et al.}~\cite{Wang14} conducted a spin pumping experiment for a trilayer of the same composition, and found that the insertion of an antiferromagnetic layer substantially enhances the spin pumping signal. These experiments not only confirmed that an antiferromagnet can transmit spins but also demonstrated that an antiferromagnet is an efficient spin conductor. 

Stimulated by these experiments, in 2016 Qiu {\it et al.} measured temperature dependence of the spin pumping in a Y$_3$Fe$_5$O$_{12}$/CoO/Pt trilayer, and found that the spin pumping signal shows a peak with its peak temperature correlated with the N\'{e}el temperature $T_{\rm N}$ of CoO layer~\cite{Qiu16}. More surprisingly, they found that the spin pumping signal strongly depends on the magnitude of external microwave frequency, a reminiscent of the dynamic critical phenomena in antiferromagnets~\cite{Neighbours63}. We note that a similar enhancement of the spin current near $T_{\rm N}$ was also observed in the spin pumping in NiFe/IrMn~\cite{Frangou16} and in the spin Seebeck effect in Y$_3$Fe$_5$O$_{12}$/NiO/Pt~\cite{Lin16}. 

In the literature, there have been several theoretical publications that attempt to explain the correlation between the peak of spin pumping signal and $T_{\rm N}$ of the antiferromagnetic layer~\cite{Okamoto16,Chen16,Reitz23}. Although they succeed in reproducing the peak structure of the spin pumping signal at $T_{\rm N}$, they do not account for another important experimental finding~\cite{Qiu16} that the spin pumping has a strong dependence on the external microwave frequency upon approaching the N\'{e}el temperature (see Fig.~4 of Ref.~\cite{Qiu16}). Given the potential of antiferromagnets as an efficient spin transmission line~\cite{Hou19}, it is of vital importance to develop a theory that can simultaneously explain the peak structure at $T_{\rm N}$ and the strong frequency dependence of the spin pumping signal found in a ferromagnet/antiferromagnet/heavy metal trilayer~\cite{Qiu16}. 

In this paper, we develop Ginzburg-Landau (GL) theory of the spin pumping in a FI/AFI/M trilayer, where FI is a ferromagnetic insulator, AFI is a fluctuating antiferromagnetic insulator near the N\'{e}el temperature $T_{\rm N}$, and M is a heavy metal. First, based on the time-dependent Ginzburg-Landau (TDGL) theory~\cite{Adachi18,Yamamoto19,Yamamoto22}, we calculate the spin conductance at the FI/AFI interface as well as at the AFI/M interface. Next, using the boundary conditions depeloved in Refs.~\cite{Chen16,SSLZhang12}, we investigate the spin transport through the FI/AFI/M trilayer. In calculating the spin conductance for the interfaces, we consider two types of the interfacial exchange interaction; the coupling between a spin $\bmS$ in the FI layer and the magnetization vector $\bmm$ in the AFI layer (we term this interaction ``magnetic coupling''), and the coupling between a spin $\bmS$ in the FI layer and the N\'{e}el vector $\bmn$ in the AFI layer (we term this interaction ``N\'{e}el coupling''). We note that the N\'{e}el coupling was proposed in the early stage of antiferromagnetic spintronics~\cite{Takei15} [see Eq.~(1) therein], and there has been increasing evidence that the N\'{e}el coupling plays an important role in explaining several spin transport experiments~\cite{Yamamoto22,Fukushima25}. With these theoretical tools, we investigate the spin pumping signal in the FI/AFI/M trilayer. Then, regarding the temperature dependence of the spin pumping signal, we find that when the FI/AFI interface is described by the magnetic coupling, a moderate cusp appears at the N\'{e}el temperature $T_{\rm N}$ with no visible frequency dependence. By contrast, when there is a sizable ``N\'{e}el coupling'' at the FI/AFI interface, a pronounced peak appears at $T_{\rm N}$ with a strong frequency dependence. We argure that the latter result is consistent with the spin pumping experiment for a Y$_3$Fe$_5$O$_{12}$/CoO/Pt trilayer~\cite{Qiu16}.

This paper is organized as follows. In Sec.~\ref{Sec:II}, we describe our model. In the subsequent three sections, we develop GL theory of the spin pumping in a FI/AFI/M trilayer. In Sec.~\ref{Sec:III} we use TDGL equation and calculate microscopically the spin conductance at the FI/AFI interface, while in Sec.~\ref{Sec:IV} we calculate the spin conductance at the AFI/M interface. In Sec.~\ref{Sec:V}, using the boundary conditions for the spin current, we discuss the spin transport in the FI/AFI/M trilayer. Then, in Sec.~\ref{Sec:VI} we compare our theoretical result to experiments, and finally, in Sec.~\ref{Sec:VII} we discuss and summarize our results.

\section{Model \label{Sec:II}  } 

We begin with the free energy of the FI layer, 
\begin{equation}
  F_{\rm F} = F_0 (|\bmS|) 
  - \gamma \hbar \bm{H}_{\text{ex}} \cdot \bmS, 
\end{equation}
where $F_0(|\bmS|)$ is the free energy at zero external magnetic field, which is independent of the direction of spin $\bmS$ and thus depends only on its magnitude, $|\bmS|$~\cite{Landau-statphys2}. The magnetic anisotropy is discarded here. The second term on the right-hand side describes the Zeeman coupling of $\bmS$ with the external magnetic field  $\bmH_{\rm ex}$, where $\gamma$ and $\hbar$ are the gyromagnetic ratio and Planck constant. The external magnetic field has two contributions,
\begin{eqnarray}
  \bmH_{\text{ex}} = \bmH_0 + \bmh_{\rm ac} (t),
\end{eqnarray}
where $\bm{H}_0 = H_0 {\bf \hat{z}}$ is a static magnetic field, and
\begin{equation}
  \bmh_{\rm ac} (t) = \frac{h_{\rm ac}}{2}
  \Big( e^{-i \omega t}+  e^{i \omega t} \Big) {\bf \hat{x}} 
  \label{eq:h_ac01}
\end{equation}
is an ac magnetic field that drives the ferromagnetic resonance. Note that, as the spin pumping driven by the ferromagnetic resonance involves only the uniform (Kittel) mode of magnons, throughout this paper we disregard spatially nonuniform magnon modes. 

Next, we consider the free energy of the AFI layer~\cite{Yamamoto19,Yamamoto22,Landau-electrodyn}, 
\begin{eqnarray}
  F_{\rm A} &=& \epsilon _0 v_0 \biggl\{ \frac{u_2}{2}\bm{n}^2 + \frac{u_4}{4}(\bm{n}^2)^2 + \frac{K_0}{2}(\bm{n}\times {\bf \hat{u}} )^2   \nonumber\\
  && 
  + \frac{D'}{2} \bm{m}^2\bm{n}^2 + \frac{r_0}{2} \bm{m}^2 \biggl\} 
  - \gamma \hbar \bmH_{\rm ex} \cdot \bmm , 
\end{eqnarray}
where $\bm{m}$ and $\bm{n}$ are respectively the magnetization vector and staggered (N\'{e}el) vector, both of which are coarse-grained within an effective cell volume $v_0$. Here, $\epsilon _0$ is the magnetic energy density, $\mathfrak{h}_0 = \epsilon _0 v_0/\gamma \hbar$ is the unit of magnetic field. As for other dimensionless coefficients, $u_2=(T-T_{\rm N})/T_{\rm N}$ is the quadratic coefficient, $u_4$ is the quartic coefficient, $r_0^{-1}=A(T/T_{\rm N}+\Theta)^{-1}$ with two parameters $A$ and $\Theta$ corresponds to the paramagnetic susceptibility, and $D'$ describes the coupling between $\bm{m}$ and $\bm{n}$~\cite{Landau-electrodyn}. Finally, $K_0$ describes the uniaxial anisotropy along the $z$ axis, therefore we take ${\bf \hat{u}} = {\bf \hat{z}}$ unless otherwise stated. 

We also consider the free energy of the M layer~\cite{Yamamoto22},  
\begin{equation}
  F_{\rm M} = \frac{1}{2 \chi_{\rm M}(0)} \bmsig^2
  - \gamma \hbar \bmH_{\rm ex}  \cdot \bmsig, 
\end{equation}
where $\bmsig$ and $\chi_{\rm M}(0)$ are respectively the itinerant spin density and the static spin susceptibility in the M layer. 

\begin{figure}[t] 
  \begin{center}
    \includegraphics[width=8cm]{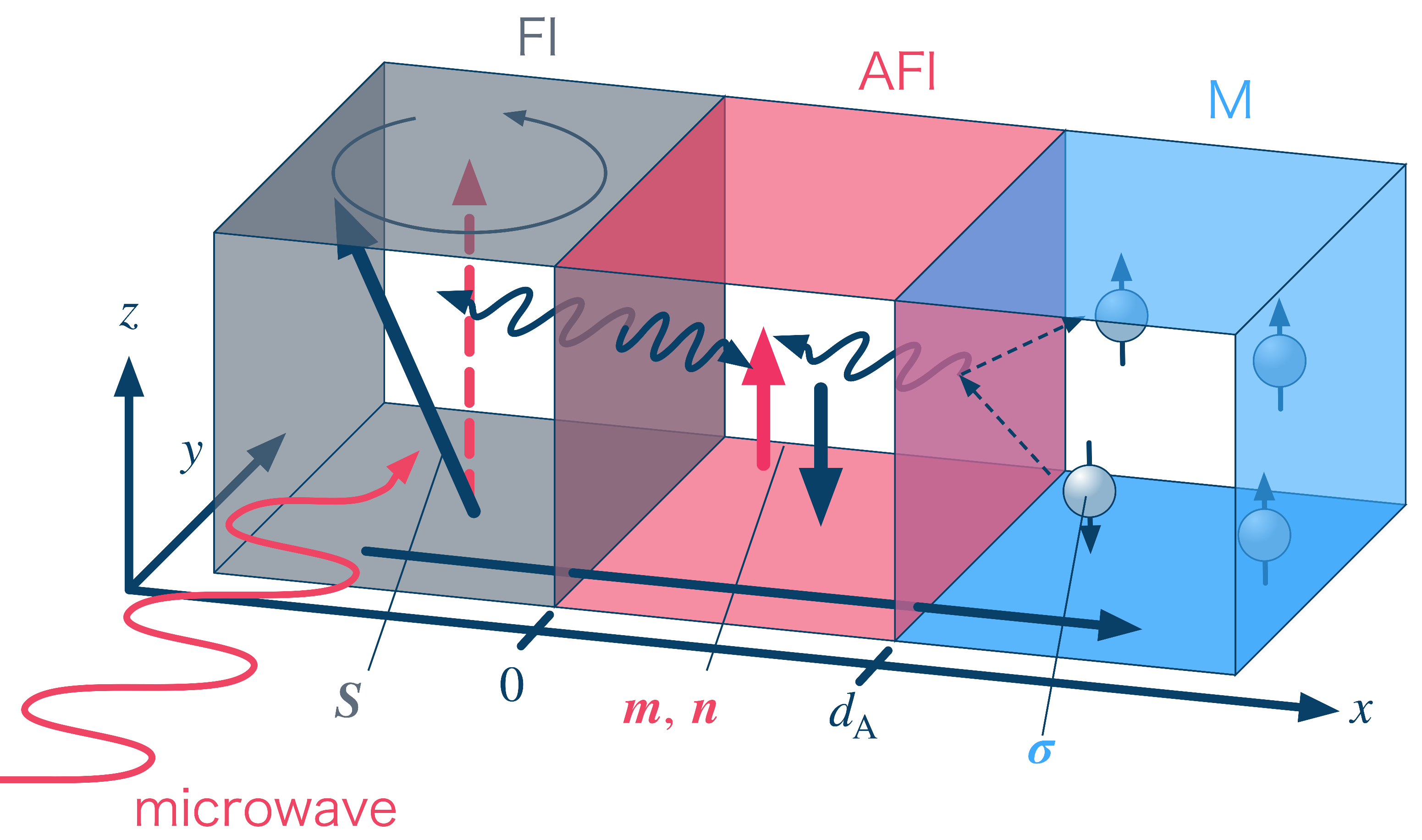} 
  \end{center}
  \caption{
    Schematic illustration of the system considered in this work. Here, FI, AFI, and M refer to a ferromagnetic insulator, antiferromagnetic insulator, and heavy metal, respectively, and $d_{\rm A}$ is the thickness of the AFI layer. Besides, $\bmS$ is the spin in the FI layer, $\bmm$ and $\bmn$ respectively represent the magnetization vector and N\'{e}el vector in the AFI layer, and $\bmsig$ denotes itinerant spin density in the M layer.
  } 
  \label{fig:schematic}
\end{figure}

Now, we focus on the interaction at the FI/AFI interface, as well as at the AFI/M interface. Following the argument of Ref.~\cite{Yamamoto22}, we consider two types of the interfacial exchange interaction. The first is the coupling between the spin $\bmS$ in the FI layer and magnetization vector $\bmm$ in the AFI layer, which we term ``magnetic coupling''. The second is the coupling between the spin $\bmS$ in the FI layer and the N\'{e}el vector $\bmn$ in the AFI layer, which we term ``N\'{e}el coupling''. Therefore, the interaction at the FI/AFI interface is described by the following free energy, 
\begin{subequations}
  \begin{empheq}[left = 
      {   F_{\rm F/A} = \empheqlbrace \,}]{alignat = 2} 
    & - J_m \bmS \cdot \bmm  &\;\;\;&   \text{for \quad magnetic~coupling}, 
    \label{eq:magn-int01}\\
    & -J_n \bmS \cdot \bmn  & &  \text{for \quad N\'{e}el~coupling},
    \label{eq:neel-int01}
    \end{empheq}
\end{subequations}
where $J_m$ ($J_n$) is the magnetic (N\'{e}el) coupling constant at the FI/AFI interface. In a similar way, at the AFI/M interface, we consider the free energy 
\begin{subequations}
  \begin{empheq}[left = 
      {   F_{\rm A/M} = \empheqlbrace \,}]{alignat = 2} 
    & - J'_m \bmm \cdot \bmsig
    &\;\;\;&   \text{for \quad magnetic~coupling}, 
    \label{eq:magn-int02}\\
    & -J'_n \bmn \cdot \bmsig  & &  \text{for \quad N\'{e}el~coupling},
    \label{eq:neel-int02}
    \end{empheq}
\end{subequations}
where $J'_m$ ($J'_n$) is the magnetic (N\'{e}el) coupling constant at the AFI/M interface. 

We now discuss the spin dynamics in FI, AFI, and M layers. The dynamics of $\bmS$ in the FI layer is described by the Landau-Lifshitz-Gilbert (LLG) equation~\cite{Landau-statphys2}, 
\begin{equation}
  \frac{\partial}{\partial t} \bmS = \gamma \bmH_{\rm eff} \times \bmS 
  + \frac{\alpha}{|\bmS|} \bmS \times \frac{\partial}{\partial t} \bmS,
  \label{eq:LLG01}
\end{equation}
where $\alpha$ is the Gilbert damping constant, and
\begin{equation}
  \bmH_{\rm eff}= -\frac{1}{\gamma \hbar} \frac{\partial}{\partial \bmS} \left( F_{\rm F} + F_{\rm F/A} \right)
\end{equation}
is the effective field for $\bmS$.

The dynamics of $\bm{m}$ and $\bm{n}$ in the AFI layer is described by the TDGL equations~\cite{Yamamoto19, Freedman76,Halperin76}: 
\begin{eqnarray}
  \label{eq:TDGL_m01}
  \frac{\partial}{\partial t} \bm{m}
  &=& \gamma \bm{H}_m \times \bm{m}+\gamma \bm{H}_n \times \bm{n}+\Gamma _m \frac{\bm{H}_m}{\mathfrak{h}_0 }, 
\\
  \label{eq:TDGL_n01}
    \frac{\partial}{\partial t} {\bm n}
  &=& \gamma {\bm H}_n \times {\bm m}+\gamma {\bm H}_m \times {\bm n}+\Gamma _n \frac{{\bm H}_n}{\mathfrak{h}_0} , 
\end{eqnarray}
where $\Gamma _m$ and $\Gamma_n$ are damping coefficients for $\bmm$ and $\bmn$, respectively, and the effective fields $\bm{H}_m$ and $\bm{H}_n$ are defined by
\begin{eqnarray}
  \bmH_m &=& -\frac{1}{\gamma \hbar} \frac{\partial}{\partial \bm{m}}
  \left( F_{\rm A}+F_{\rm int} \right), \\
  \bmH_n &=& -\frac{1}{\gamma \hbar} \frac{\partial}{\partial \bmn}
  \left( F_{\rm A} +F_{\rm int} \right), 
\end{eqnarray}
where $F_{\rm int} = F_{\rm F/A} $ when we discuss the spin conductance at the FI/AFI interface, whereas $F_{\rm int} = F_{\rm A/M} $ when we discuss the spin conductance at the AFI/M interface.

Finally, in the M layer, the dynamics of $\bmsig$ is described by the Bloch equation, 
\begin{equation}
  \frac{\partial}{\partial t} \bmsig
  = \gamma \bmH_{\sigma} \times \bmsig
  + \frac{\chi_{\rm M}(0) \gamma \hbar}{\tau_{\rm M}} \bmH_{\sigma} ,
  \label{eq:Bloch01}
\end{equation}
where $\tau_{\rm M}$ is the spin-flip relaxation time of $\bmsig$, and 
\begin{equation}
  \bmH_{\sigma} = -\frac{1}{\gamma \hbar} \frac{\partial}{\partial \bmsig}
  \left( F_{\rm M} + F_{\rm A/M} \right)  
\end{equation}
is the effective field for $\bmsig$. 

Before ending this section, we discuss the equilibrium spin configuration in the FI and AFI layers. In the FI layer, because we assume that the spins are fully saturated, the equilibrium spin configuration is given by  
\begin{equation}
  \bmS_{\rm eq} = S_0 {\bf \hat{z}},
  \label{eq:Seq01}
\end{equation}
where $S_0$ is determined by minimizing the free energy $F_0$, i.e., $F_0'(S_0)=0$~\cite{Landau-statphys2}. In the AFI layer, the equilibrium value of the order parameter (N\'{e}el vector) is determined by the condition $\bmH_n = {\bm 0}$, from which we obtain
\begin{equation}
  \bmn_{\rm eq} = n_{\rm eq} {\bf \hat{z}},
  \label{eq:Neq01}
\end{equation}
where 
\begin{equation}
  n_{\rm eq} =
  \begin{cases}
    \sqrt{\frac{|u_2|}{u_4}} = \sqrt{\frac{T_{\rm N}-T}{T_{\rm N}} \frac{1}{u_4}} & (T \leq T_{\rm N}), \\
    0 & (T>T_{\rm N}). 
  \end{cases}
\end{equation}

In the AFI layer, the equilibrium value of 
\begin{equation}
  \bmm_{\rm eq} = m_{\rm eq} {\bf \hat{z}} 
  \label{eq:Meq01}  
\end{equation}
is determined by the condition $\bmH_m= {\bm 0}$, from which we obtain 
\begin{equation}
  m_{\rm eq} = {\chi}_{\rm A}(0) \Big( g \mu_{\rm B} {H_0} + J_m S_0 \Big), 
\end{equation}
where $g$ is the g-factor, $\mu_{\rm B}$ is the Bohr magneton, and 
\begin{equation}
  {\chi}_{\rm A}(0) = \frac{1}{\epsilon_0 v_0 (r_0 + D' n_{\rm eq}^2)}
\end{equation}
is the static spin susceptibility of the AFI layer. Note that in deriving the above results, we assumed $m_{\rm eq} \ll n_{\rm eq}$.

\section{Spin conductance at FI/AFI interface \label{Sec:III}}  

  In the subsequent three sections, we develop GL theory of the spin pumping in the FI/AFI/M trilayer as shown in Fig.~\ref{fig:schematic}. The procedure consists of three steps. First, in Sec.~\ref{Sec:III}, we use the TDGL theory and calculate the spin conductance at the FI/AFI interface. Next, in Sec.~\ref{Sec:IV}, we calculate the spin conductance at the AFI/M interface by the TDGL theory. Finally, in Sec.~\ref{Sec:V}, we use the boundary conditions developed in Refs.~\cite{Chen16,SSLZhang12} and calculate the spin transport in the FI/AFI/M trilayer.

Now, we move on to the first step and calculate the spin conductance at the FI/AFI interface. For this purpose, we consider the spin pumping from the FI into the AFI layer, assuming that the interaction at the FI/AFI interface is described either by the magnetic coupling [Eq.~(\ref{eq:magn-int01})] or the N\'{e}el coupling [Eq.~(\ref{eq:neel-int01})]. In the following analysis, we consider non-equilibrium fluctuations of $\bmS$, $\bmn$, and $\bmm$ by introducing the following decomposition:
\begin{eqnarray}
  \bmS = \bmS_{\rm eq} + \delta \bmS, \label{eq:SeqdS01}\\
  \bmn = \bmn_{\rm eq} + \delta \bmn, \label{eq:NeqdN01}\\
  \bmm = \bmm_{\rm eq} + \delta \bmm. \label{eq:MeqdM01}
\end{eqnarray}
The spin current density pumped from the FI into the AFI layer is defined by~\cite{Adachi13,Fukushima25} 
\begin{equation}
  j_{\rm s}(t) = -\frac{1}{{\cal S}_{\rm F/A}} \frac{\partial}{\partial t} m^z(t) \Big]_{\rm interface},   \label{eq:js01}
\end{equation}
where ${\cal S}_{\rm F/A}$ is the interface cross area, and $ \cdots ]_{\rm interface}$ means to specify a rate of change due to the interfacial spin transfer. Note that a magnon carry spin $-\hbar$, and that we take the spin quantization axis along the $-z$ direction so that the sign of the pumped spin current is consistent with other literature~\cite{Tserkovnyak02,Chen16}.

\subsection{Magnetic coupling} 
When the interaction at the FI/AFI interface is described by the magnetic coupling [Eq.~(\ref{eq:magn-int01})], using the $z$ component of the TDGL equation (\ref{eq:TDGL_m01}) we obtain~\cite{Adachi13,Adachi18}
\begin{equation}
  j_{\rm s} (t) = \frac{J_m}{{\cal S}_{\rm F/A} \hbar } \operatorname{Im} [ \delta S^+(t) \delta m^-(t)],
  \label{eq:js02}
\end{equation}
where we define $\delta S^\pm = \delta S^x \pm i \delta S^y$ and $\delta m^\pm = \delta m^x \pm i \delta m^y$. Because we are considering a monochromatic disturbance $\bmh_{\rm ac}$ [see Eq.~(\ref{eq:h_ac01})], the corresponding linearized solutions are generally expressed as~\cite{Landau-statphys1}
\begin{eqnarray}
  \delta S^+(t) &=& \delta S^+_{\omega} e^{- i \omega t} + \delta S^+_{-\omega} e^{i \omega t}, \label{eq:dSpls01}\\
  \delta m^-(t) &=& \delta m^-_{\omega} e^{- i \omega t} + \delta m^-_{-\omega} e^{i \omega t}. \label{eq:dMmns01}
\end{eqnarray}
Then, substituting Eqs.~(\ref{eq:dSpls01}) and (\ref{eq:dMmns01}) into Eq.~(\ref{eq:js02}) and focusing only on the dc component, we obtain 
\begin{equation}
  j_{\rm s} = \frac{J_m}{{\cal S}_{\rm F/A} \hbar } \operatorname{Im}
  [ \delta S^+_{-\omega} \delta m^-_{\omega}] + (\omega \rightarrow -\omega). 
  \label{eq:js03}
\end{equation}
Therefore, our remaining task is to calculate $\delta S^+_{-\omega}$ and $\delta m^-_{\omega}$ by using the LLG equation (\ref{eq:LLG01}) and TDGL equations (\ref{eq:TDGL_m01}) and (\ref{eq:TDGL_n01}). 

We first calculate  $\delta S^+_{-\omega}$. Linearizing the LLG equation~(\ref{eq:LLG01}) with respect to $\delta \bmS_\omega$ and projecting $\delta \bmS_\omega$ onto $\delta S_\omega^\pm$, we obtain 
\begin{eqnarray}
  \delta S_{\omega}^+ &=& \chi_{\rm F} (-\omega)^* 
  \Big( \frac{\hbar \gamma h_{\rm ac}}{2} + {J_m} \delta m^+_{\omega} \Big), \label{eq:dSp01} \\
  \delta S_{\omega}^- &=& \chi_{\rm F} (\omega) 
  \Big( \frac{\hbar \gamma h_{\rm ac}}{2} + {J_m} \delta m^-_{\omega} \Big), \label{eq:dSm01} 
\end{eqnarray}
where
\begin{equation}
  \chi_{\rm F} (\omega) = \frac{-S_0/\hbar}{\omega- \gamma H_0 + i \alpha \omega}  
\end{equation}
is the dynamic spin susceptibility of the FI layer.

We next calculate $\delta m^-_{\omega}$. Linearizing the TDGL equations~(\ref{eq:TDGL_m01}) and (\ref{eq:TDGL_n01}) with respect to $\delta \bmm_\omega$ and $\delta \bmn_\omega$, projecting $\delta \bmm_\omega$ and $\delta \bmn_\omega$ onto $\delta m_\omega^-$ and $\delta n_\omega^-$, we obtain 
\begin{equation}
(\omega-\widehat{\mathcal{A}}) 
\begin{pmatrix}
\delta m^{-}_\omega \\
\delta n^{-}_\omega \\
\end{pmatrix}
=
-\frac{J_m}{\hbar}
\begin{pmatrix}
m_{\rm eq}- i \frac{\Gamma_m}{\gamma \mathfrak{h}_0} \\
n_{\rm eq} \\
\end{pmatrix} \delta S^-_{\omega},
\label{eq:TDGL_matrixAF01}
\end{equation}
where the matrix $\widehat{\mathcal{A}}$ is given by 
\begin{equation}
  \widehat{\mathcal{A}} = 
\begin{pmatrix}
  a & b\\
  c & d \\
\end{pmatrix}, 
\end{equation}
with 
\begin{eqnarray}
  a &=& \gamma H_0- i \Gamma_m/\epsilon_0 v_0 {\chi}_{\rm A}(0), \\
  b &=& \gamma \mathfrak{h}_0 K n_{\rm eq}, \\
  c &=& {\gamma \mathfrak{h}_0 n_{\rm eq}}/\epsilon_0 v_0 {\chi}_{\rm A}(0), \\
  d &=& \gamma \mathfrak{h}_0 Km_{\rm eq} - i \Gamma_nK,
\end{eqnarray}
where $K= K_0 + u_2 + u_4 n_{\rm eq}^2$. Note that $K= K_0$ at $T \leq T_{\rm N}$ whereas $K= K_0 + u_2$ at $T > T_{\rm N}$. In the following calculation, it is convenient to introduce the propagator $\widehat{\mathcal{G}}=(\omega - \widehat{\mathcal{A}})^{-1}$, where each component of $\widehat{\mathcal{G}}$ is given by
\begin{eqnarray}
  \widehat{\mathcal{G}} (\omega) &=& 
  \begin{pmatrix}
    \mathcal{G}_1 (\omega) & \mathcal{G}_2 (\omega) \\
    \mathcal{G}_3 (\omega) & \mathcal{G}_4 (\omega) \\
  \end{pmatrix} \nonumber \\
  &=&
  \frac{1}{\det   (\omega- \widehat{\mathcal{A}}) }
  \begin{pmatrix}
    \omega- d  & b\\
    c & \omega - a \\
  \end{pmatrix}.
  \label{eq:propagator01}
\end{eqnarray}
Using this propagator, Eq.~(\ref{eq:TDGL_matrixAF01}) is solved for $\delta m^{-}_\omega$ as 
\begin{eqnarray}
  \delta m^{-}_\omega &=& {\chi}_{\rm A} (\omega) 
  {J_m } \delta S^-_\omega \nonumber \\
  &=&
  J_m   {\chi}_{\rm A} (\omega) 
  \chi_{\rm F} (\omega)
  \Big( \frac{\hbar \gamma h_{\rm ac}}{2} + {J_m} \delta m^-_\omega \Big),
  \label{eq:dMmns02}
\end{eqnarray}
where we used Eq.~(\ref{eq:dSm01}) in moving to the second line, and 
\begin{equation}
  {\chi}_{\rm A} (\omega)
  = - {\chi}_{\rm A} (0)
  \Big( \mathcal{G}_1 (\omega) a + \mathcal{G}_2 (\omega) c \Big) 
\end{equation}
is the dynamic spin susceptibility of the AFI layer.

Substituting Eqs.~(\ref{eq:dSp01}) and (\ref{eq:dMmns02}) into Eq.~(\ref{eq:js03}), and then summarize the result up to ${\cal O}(J_m^2)$, we finally obtain
\begin{equation}
  j_{\rm s} = \frac{J_m^2 }{{\cal S}_{\rm F/A} \hbar }
  |\chi_{\rm F}(\omega)|^2 \operatorname{Im} {\chi}_{\rm A} (\omega)
  \left( \frac{ \hbar \gamma h_{\rm ac}}{2} \right)^2  + (\omega \rightarrow -\omega),
  \label{eq:js04}  
\end{equation}
where we pick up the most dominant term under the resonance condition $\omega= \gamma H_0$ in the $\alpha \ll 1$ limit.

The above result can be summarized by using the ``spin pumping battery'' concept~\cite{Brataas02} and interface spin conductance~\cite{Chen16,SSLZhang12}. We first consider the quantity, 
\begin{equation}
  {\bf \hat{z}} \cdot (\bmS \times \partial_t \bmS )
  =
  \operatorname{Im} \Big[ \delta S^-(t) \; \partial_t \delta S^+(t)  \Big], 
\end{equation}
where the dynamics of $\delta \bmS(t)$ is evaluated in the zeroth order with respect to the interface exchange interactions. Then, substituting Eqs.~(\ref{eq:dSpls01}), (\ref{eq:dSp01}), and (\ref{eq:dSm01}) into the above equation and focusing only on the dc component, we obtain 
\begin{eqnarray}
  {\bf \hat{z}} \cdot (\bmS \times \partial_t \bmS )
  &=& \omega \, \delta S^-_\omega \delta S^+_{-\omega}
  + (\omega \rightarrow -\omega) \hspace{2cm} \nonumber \\
  &=& \omega |\chi_{\rm F}(\omega)|^2
  \left( \frac{\hbar \gamma h_{\rm ac}}{2} \right)^2 + (\omega \rightarrow -\omega).
  \label{eq:SxdS01}
\end{eqnarray}
Using Eq.~(\ref{eq:SxdS01}) and recalling that $\operatorname{Im} {\chi}_{\rm A} (\omega)$ is linear in $\omega$, the spin current pumped from the FI to the AFI layer is represented as 
\begin{equation}
    j_{\rm s} = g_{\rm F/A} (\omega) \;  V_{\rm s}, 
  \label{eq:js_gFA01}
\end{equation}
where
\begin{equation}
  V_{\rm s} = {\bf \hat{z}} \cdot (\bmS \times \partial_t \bmS )
  \label{eq:Vspin01}
\end{equation}
is the spin voltage generated by the spin pumping battery, and  
\begin{equation}
  g_{\rm F/A} (\omega) = \frac{J_m^2}{\hbar {\cal S}_{\rm F/A}} 
  \frac{1}{\omega} \operatorname{Im} {\chi}_{\rm A} (\omega)
  \label{eq:gFA01}
\end{equation}
is the spin conductance at the FI/AFI interface for the present magnetic coupling. Note that ${\omega}^{-1} \operatorname{Im} {\chi}_{\rm A} (\omega)$ can be expressed in terms of ${\cal G}_1$ and ${\cal G}_2$ as
\begin{equation}
  \frac{1}{\omega}   \operatorname{Im} {\chi}_{\rm A} (\omega) 
  = \frac{1}{\epsilon_0 v_0} \Big( \Gamma_m |{\cal G}_1 (\omega)|^2
  + \Gamma_n |{\cal G}_2 (\omega)|^2 \Big). 
  \label{eq:ImCHI01}
\end{equation}

\subsection{N\'{e}el coupling}

When the interaction at the FI/AFI interface is described by the N\'{e}el coupling [Eq.~(\ref{eq:neel-int01})], the exchange field $(J_n/\gamma \hbar) \bmS$ at the FI layer exerts the torque to the N\'{e}el vector $\bmn$ in the AFI layer. Thus, the spin current pumped into the AFI layer is not given by Eq.~(\ref{eq:js02}), but instead given by 
\begin{equation}
  j_{\rm s} = \frac{J_n}{{\cal S}_{\rm F/A} \hbar } \operatorname{Im} [ \delta S^+(t) \delta n^-(t)],  
  \label{eq:js05}
\end{equation}
where $\delta n^\pm = \delta n^x \pm i \delta n^y$. As in the magnetic coupling case, because of the monochromatic nature of the disturbance [see Eq.~(\ref{eq:h_ac01})], the corresponding linearized solutions are generally given by Eq.~(\ref{eq:dSpls01}) and  
\begin{equation}
  \delta n^-(t) = \delta n^-_{\omega} e^{- i \omega t} + \delta n^-_{-\omega} e^{i \omega t}. \label{eq:dNmns01}
\end{equation}
Then, substituting Eqs.~(\ref{eq:dSpls01}) and (\ref{eq:dNmns01}) into Eq.~(\ref{eq:js05}) and focusing only on the dc component, we obtain 
\begin{equation}
  j_{\rm s} = \frac{J_n}{{\cal S}_{\rm F/A} \hbar } \operatorname{Im}
  [ \delta S^+_{-\omega} \delta n^-_{\omega}] + (\omega \rightarrow -\omega). 
  \label{eq:js06}
\end{equation}

Now, the rest of the calculation is formally quite similar to that of the magnetic coupling case. In the present N\'{e}el coupling case, the fluctuation $\delta S^\pm_{\omega}$ is quite similar to Eqs.~(\ref{eq:dSp01}) and (\ref{eq:dSm01}) as  
\begin{eqnarray}
  \delta S_{\omega}^+ &=& \chi_{\rm F} (-\omega)^*
  \Big( \frac{\hbar \gamma h_{\rm ac}}{2} + {J_n} \delta n^+_{\omega} \Big), 
  \label{eq:dSp02} \\
  \delta S_{\omega}^- &=& \chi_{\rm F} (\omega) 
  \Big( \frac{\hbar \gamma h_{\rm ac}}{2} + {J_n} \delta n^-_{\omega} \Big), 
  \label{eq:dSm02}
\end{eqnarray}
but calculation of $\delta n^-_{\omega}$ requires some care. For the N\'{e}el coupling, the TDGL equation for $\delta m^-_\omega$ and $\delta n^-_\omega$ is given by 
\begin{equation}
(\omega-\widehat{\mathcal{A}}) 
\begin{pmatrix}
\delta m^{-}_\omega \\
\delta n^{-}_\omega \\
\end{pmatrix}
=
-\frac{J_n}{\hbar}
\begin{pmatrix}
n_{\rm eq} \\  
m_{\rm eq}- i \frac{\Gamma_n}{\gamma \mathfrak{h}_0} \\
\end{pmatrix} \delta S^-_{\omega},
\label{eq:TDGL_matrixAF02}
\end{equation}
and solving this equation by using the propagator [Eq.~(\ref{eq:propagator01})], the fluctuation $\delta n^-_\omega$ is given by
\begin{eqnarray}
  \delta n^{-}_\omega &=& {\psi}_{\rm A} (\omega) 
  {J_n } \delta S^-_\omega \nonumber \\
  &=&
  {J_n }  
  {\psi}_{\rm A} (\omega) 
  \chi_{\rm F} (\omega)
  \Big( \frac{ \hbar \gamma h_{\rm ac}}{2} + {J_n} \delta n^-_{\omega} \Big),
  \label{eq:dNmns02}  
\end{eqnarray}
where we used Eq.~(\ref{eq:dSm02}) in moving to the second line.
In the above equation, 
\begin{equation}
  {\psi}_{\rm A} (\omega)
  = - {\psi}_{\rm A} (0) 
  \Big( \mathcal{G}_3 (\omega) b + \mathcal{G}_4 (\omega) d \Big) 
\end{equation}
is the dynamic N\'{e}el susceptibility of the AFI layer, where 
\begin{equation}
  {\psi}_{\rm A}(0) = \frac{1}{\epsilon_0 v_0 K}  
\end{equation}
is the static N\'{e}el susceptibility.

Now, substituting Eqs.~(\ref{eq:dSp02}) and (\ref{eq:dNmns02}) into Eq.(\ref{eq:js06}) and summarizing the result up to ${\cal O}(J_n^2)$, we finally obtain 
\begin{equation}
  j_{\rm s} = \frac{J_n^2 }{{\cal S}_{\rm F/A} \hbar} 
  |\chi_{\rm F}(\omega)|^2 \operatorname{Im} {\psi}_{\rm A} (\omega)
  \left( \frac{\hbar \gamma h_{\rm ac}}{2} \right)^2 + (\omega \rightarrow -\omega). 
  \label{eq:js07}
\end{equation}
Using Eq.~(\ref{eq:SxdS01}), the above result can be summarized in the same form as Eq.~(\ref{eq:js_gFA01}). In the present N\'{e}el coupling case, the spin conductance at the FI/AFI interface is given by 
\begin{equation}
  g_{\rm F/A} (\omega) = \frac{J_n^2}{\hbar {\cal S}_{\rm F/A}} 
  \frac{1}{\omega} \operatorname{Im} {\psi}_{\rm A} (\omega), 
  \label{eq:gFA02}
\end{equation}
where $\omega^{-1} \operatorname{Im}\psi_{\rm A} (\omega)$ can be expressed in terms of ${\cal G}_3$ and ${\cal G}_4$ as 
\begin{equation}
  \frac{1}{\omega}  \operatorname{Im} {\psi}_{\rm A} (\omega)
  = \frac{1}{\epsilon_0 v_0 } \Big( 
  \Gamma_n |{\cal G}_4 (\omega)|^2 
  + \Gamma_m |{\cal G}_3 (\omega)|^2 \Big).
  \label{eq:ImPSI01}  
\end{equation}

\section{Spin conductance at AFI/M interface  \label{Sec:IV} }
In this section, we examine the spin conductance at the AFI/M interface. To this end, we consider the spin injection that is driven by a non-equilibrium magnon accumulation created at the AFI layer. Technically speaking, the corresponding calculation of the spin conductance is much more involved than that presented in the previous section. This is because the spin transfer across the AFI/M interface is driven by a non-equilibrium magnon accumulation, and its description inevitably requires us to consider the deviation of magnon distribution function from its thermal equilibrium value. This means that, within the present model, we need to include thermal noise fields ${\bm \xi}$ and ${\bm \eta}$ at the right-hand side of the TDGL Eqs.~(\ref{eq:TDGL_m01}) and (\ref{eq:TDGL_n01}) for the AFI layer, each of which has zero mean and variance~\cite{Adachi18,Yamamoto19,Yamamoto22}, 
\begin{eqnarray}
  \bra \xi^i (t) \xi^j (t') \ket &=& 
  \frac{2 \kB T_{\rm A} \Gamma_m}{\epsilon_0 v_0} \delta_{i,j} \delta(t-t'), \\
    \bra \eta^i (t) \eta^j (t') \ket &=& 
  \frac{2 \kB T_{\rm A} \Gamma_n}{\epsilon_0 v_0} \delta_{i,j} \delta(t-t'), 
\end{eqnarray}
where $T_{\rm A}$ is the temperature of the AFI layer, $\bra \cdots \ket$ represents averaging over thermal noise, $i$ and $j$ denote $x,y,z$, and the noise fields satisfy 
\begin{equation}
  \bra \xi^i (t) \eta^j (t') \ket = 0, 
\end{equation}
which means that ${\bm \xi}$ and ${\bm \eta}$ are independent. 

Let us turn to the dynamics of the M layer. As in Eqs.~(\ref{eq:SeqdS01}), (\ref{eq:NeqdN01}), and (\ref{eq:MeqdM01}), we introduce the decomposition 
\begin{equation}
  \bmsig = \bmsig_{\rm eq} + \delta \bmsig, \label{eq:SIGeqdSIG01}
\end{equation}
where $\bmsig_{\rm eq}$ is the equilibrium value and $\delta \bmsig$ describes the spin accumulation. Here, $\bmsig_{\rm eq}$ is given by 
\begin{subequations}
  \begin{empheq}[left = 
      { \bmsig_{\rm eq}  = \empheqlbrace \,}]{alignat = 2} 
    & J'_{m} \chi_{\rm M}(0) \bmm_{\rm eq}
    &\;\;\;&   \text{for \quad magnetic~coupling}, \\
    & J'_{n} \chi_{\rm M}(0) \bmn_{\rm eq}    
    & &  \text{for \quad N\'{e}el~coupling}, 
    \end{empheq}
\end{subequations}
where we discarded the Zeeman term. At the right-hand side of the Bloch Eq.~(\ref{eq:Bloch01}), if we include thermal noise field ${\bm \zeta}$ with zero mean and variance 
\begin{equation}
  \bra \zeta^i (t) \zeta^j (t') \ket = 
  \frac{2 \kB T_{\rm M} \chi_{\rm M}(0)}{\tau_{\rm M}} \delta_{i,j} \delta(t-t'), 
\end{equation}
then the model becomes exactly the same as that used for the antiferromagnetic spin Seebeck effect in Refs.~\cite{Yamamoto19} and \cite{Yamamoto22}, where $T_{\rm M}$ is the temperature of the M layer. Therefore we employ the results obtained in these two works. In Refs.~\cite{Yamamoto19} and \cite{Yamamoto22}, the spin current density $j_{\rm s}$ injected into the M layer was obtained as 
\begin{equation}
  j_{\rm s} = j_{\rm s}^{\rm pump} - j_{\rm s}^{\rm back}, 
\end{equation}
where
\begin{equation}
  j_{\rm s}^{\rm pump} = 2 J'^{2}_{m/n} \frac{\kB \chi_{\rm M}(0) m_{\rm eq} \tau_{\rm M}}
  {{\cal S}_{\rm A/M} \hbar^2} T_{\rm A} 
\end{equation}
is the pumping current density, and 
\begin{equation}
  j_{\rm s}^{\rm back} = 2 J'^{2}_{m/n} \frac{\kB \chi_{\rm M}(0) m_{\rm eq} \tau_{\rm M}}
  {{\cal S}_{\rm A/M} \hbar^2} T_{\rm M} 
\end{equation}
is the backflow current density, and we introduce the following shorthand notation 
\begin{subequations}
  \begin{empheq}[left = 
      {   J'_{m/n} = \empheqlbrace \,}]{alignat = 2} 
    & J'_m 
    &\;\;\;&   \text{for \quad magnetic~coupling}, \\
    & J'_n  & &  \text{for \quad N\'{e}el~coupling}. 
    \end{empheq}
\end{subequations}
Here, we note the sign convention of the spin current mentioned below Eq.~(\ref{eq:js01}). In the above equations, ${\cal S}_{\rm A/M}$ is the AFI/M interface cross area. 

In the absence of the temperature bias ($T_{\rm A}= T_{\rm M}= T$) but the presence of the non-equilibrium magnon accumulation $\delta m^z$, the pumping current density is changed into
\begin{equation}
  j_{\rm s}^{\rm pump}
  = 2 J'^{2}_{m/n} \frac{\kB T \chi_{\rm M}(0) \tau_{\rm M}} 
  {{\cal S}_{\rm A/M} \hbar^2} \big( m_{\rm eq} + \delta m^z \big)   . 
\end{equation}
Similarly, in the absence of the temperature bias ($T_{\rm A}= T_{\rm M}= T$) but the presence of the non-equilibrium spin accumulation $\delta \sigma^z$, the backflow current density is given by 
\begin{equation}
  j_{\rm s}^{\rm back} = 2 J'^{2}_{m/n} \frac{\kB T \chi_{\rm M}(0) \tau_{\rm M}}
  {{\cal S}_{\rm A/M} \hbar^2 } m_{\rm eq} 
  + \delta j_{\rm s}^{\rm back}, 
\end{equation}
where $\delta j_{\rm s}^{\rm back}$ is given by (see Appendix~\ref{Sec:App01} for details) 
\begin{equation}
  \delta j_{\rm s}^{\rm back} =
  2J'^{2}_{m/n}  \frac{\kB T \phi_{\rm A}(0) \tau_{\rm M}}
  {{\cal S}_{\rm A/M} \hbar^2} \delta \sigma^z ,
  \label{eq:delta_js01}
\end{equation}
and $\phi_{\rm A}(0)$ is defined by  
\begin{subequations}
  \begin{empheq}[left = 
      { \phi_{\rm A}(0) = \empheqlbrace \,}]{alignat = 2} 
    & {\chi}_{\rm A}(0) 
    &\;\;\;&   \text{for \quad magnetic~coupling}, 
    \label{eq:jback_magn01}\\
    & {\psi}_{\rm A}(0) 
    & &  \text{for \quad N\'{e}el~coupling}.
    \label{eq:jback_neel01}
    \end{empheq}
\end{subequations}
Therefore, the resultant spin current density across the AFI/M interface is expressed as 
\begin{equation}
  j_{\rm s} = G_{\rm A/M} \; \delta m^z -  G'_{\rm A/M} \; \delta \sigma^z 
\end{equation}
where
\begin{equation}
  G_{\rm A/M} =
  2 J'^{2}_{m/n} \frac{\kB T \chi_{\rm M}(0) \tau_{\rm M}}
  {{\cal S}_{\rm A/M} \hbar^2}
\end{equation}
and 
\begin{equation}
  G'_{\rm A/M} =
    2 J'^{2}_{m/n} \frac{\kB T \phi_{\rm A}(0) \tau_{\rm M}}
  {{\cal S}_{\rm A/M} \hbar^2}. 
\end{equation}

\section{Spin transport in FI/AFI/M trilayer \label{Sec:V}}  
In this section, we employ the formulation of multilayer spin transport developed in Refs.~\cite{Chen16,SSLZhang12}, and investigate the spin transport in the FI/AFI/M trilayer shown in Fig.~\ref{fig:schematic}. The spin current density inside the FI layer is described by
\begin{equation}
  j_{\rm s}^{\rm F} (x) = g_{\rm F/A} (\omega) V_{\rm s} - D_{\rm F} \nabla_x \left( \frac{\delta S^z (x)}{v_0} \right),
  \label{eq:BC01}
\end{equation}
where $D_{\rm F}$ is the diffusion coefficient of the FI layer, $\delta S^z$ is the $z$ component of $\delta \bmS$ in Eq.~(\ref{eq:SeqdS01}), $g_{\rm F/A}(\omega)$ is defined in Eqs.~(\ref{eq:gFA01}) and (\ref{eq:gFA02}), and $V_{\rm s} $ is the spin voltage generated by the spin pumping battery~\cite{Brataas02} [see Eq.~(\ref{eq:Vspin01})]. In a similar way, the spin current densities inside the AFI and M layers are described by
\begin{eqnarray}
  j_{\rm s}^{\rm A} (x) &=& - D_{\rm A} \nabla_x \left( \frac{\delta m^z (x)}{v_0} \right), \\
  j_{\rm s}^{\rm M} (x) &=& - D_{\rm M} \nabla_x \left( \frac{\delta \sigma^z (x) }{v_0} \right), 
\end{eqnarray}
where $D_{\rm A}$ and $D_{\rm M}$ are the diffusion coefficients of the AFI and M layers, respectively, and $\delta m^z$ and $\delta \sigma^z$ are the $z$ components of $\delta \bmm$ [Eq.~(\ref{eq:MeqdM01})] and $\delta \bmsig$ [Eq.~(\ref{eq:SIGeqdSIG01})].

Spatial distributions of $\delta S^z$, $\delta m^z$, and $\delta \sigma^z$ are determined by the boundary conditions. At the FI/AFI interface, it is given by 
\begin{equation}
  j_{\rm s}^{\rm F}(0) = j_{\rm s}^{\rm A}(0) = G_{\rm F/A} \; \delta S^z(0)
  - G'_{\rm F/A} \; \delta m^z (0), 
  \label{eq:BC01}
\end{equation}
whereas at the AFI/M interface, 
\begin{equation}
  j_{\rm s}^{\rm A}(d_{\rm A}) = j_{\rm s}^{\rm M}(d_{\rm A})
  =  G_{\rm A/M} \delta m^z (d_{\rm A})
    - G'_{\rm A/M} \delta \sigma^z (d_{\rm A}), 
  \label{eq:BC02}  
\end{equation}
where $d_{\rm A}$ is the thickness of the AFI layer. Note that, although $G_{\rm A/M}$ and $G'_{\rm A/M}$ were calculated in the previous section, $G_{\rm F/A}$ and $G'_{\rm F/A}$ are not. This is reasoned by the following argument. First, we recall that $G_{\rm F/A}, G'_{\rm F/A} \propto J_m^2$ ($J_n^2$) and $G_{\rm A/M}, G'_{\rm A/M} \propto J'^{2}_m$ ($J'^{2}_n$) for the magnetic coupling (N\'{e}el coupling). Next, because of the difference in the size of the moment, i.e., a giant moment in the FI layer and a small moment in the M layer, we anticipate the relation $J^2_m \gg J'^{2}_{m} $ and $J^2_n \gg J'^{2}_{n}$. Then, as in Ref.~\cite{Chen16} [see the paragraph above Eq.~(27) therein], we may assume the limit $G_{\rm F/A}, G'_{\rm F/A} \gg G_{\rm A/M}, G'_{\rm A/M}$. Now, with additional condition $d_{\rm A} \ll \lambda_{\rm A} $ with $\lambda_{\rm A}$ being the spin diffusion length of the AFI layer, the final result [Eq.~(\ref{eq:jsRES01})] becomes independent of $G_{\rm F/A}$ and $G'_{\rm F/A}$~\cite{Chen16}, such that microscopic expressions of $G_{\rm F/A}$ and $G'_{\rm F/A}$ are not needed for the present purpose. Of course, this is merely a theoretical assumption and its validity should be determined by comparison with experiments.

\begin{figure}[t] 
  \begin{center}
    \includegraphics[width=8.5cm]{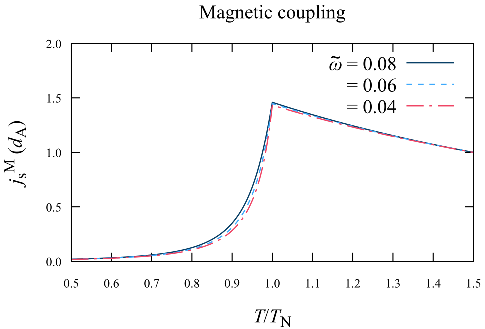} 
  \end{center}
  \caption{
    Temperature dependence of the pumped spin current density $j_{\rm s}^{\rm M}(d_{\rm A})$ for the magnetic coupling case calculated from Eq.~(\ref{eq:jsRES01}) for several values of $\widetilde{\omega}$. Here, the data is normalized by its value at $T= 1.5 T_{\rm N}$, and $\widetilde{\Gamma}_m= 0.1$, $\widetilde{\Gamma}_n= 0.5$, and $K_0= 0.1$ are used. 
  }
  \label{fig:jm-temp01} 
\end{figure}

We determine the spatial distributions of $\delta S^z$, $\delta m^z$, and $\delta \sigma^z$ by setting
\begin{eqnarray}
  \delta S^z (x) &=& C_1 e^{x/\lambda_{\rm F}}, \\
  \delta m^z (x) &=& C_2 e^{x/\lambda_{\rm A}} + C_3 e^{-x/\lambda_{\rm A}}, \\
  \delta \sigma^z (x) &=& C_4 e^{-(x- d_{\rm A})/\lambda_{\rm M}} , 
\end{eqnarray}
where $\lambda_{\rm F}$, $\lambda_{\rm A}$, $\lambda_{\rm M}$ are the spin diffusion lengths in each layer. Then, using the boundary conditions in Eqs~(\ref{eq:BC01}) and (\ref{eq:BC02}) and assuming $d_{\rm A} \ll \lambda_{\rm A} $ and $G_{\rm F/A}, G'_{\rm F/A} \gg G_{\rm A/M}, G'_{\rm A/M}$, the spin current density in the M layer is calculated to be~\cite{Chen16} 
\[
j_{\rm s}^{\rm M} (x) = \frac{g_{\rm F/A} (\omega) V_{\rm s} e^{-(x-d_{\rm A})/\lambda_{\rm M}}}
{ {1+ (1 +  {G'_{\rm A/M}}/{G_{\rm M}}){G_{\rm F}}/{G_{\rm A/M}}   } }, 
\]
where $G_{\rm F}= D_{\rm F}/v_0 \lambda_{\rm F}$, $G_{\rm A}= D_{\rm A}/v_0 \lambda_{\rm A}$, $G_{\rm M}= D_{\rm M}/v_0\lambda_{\rm M}$, and we assumed $G_{\rm F/A} \approx G'_{\rm F/A}$ for simplicity. Then, assuming $G'_{\rm A/M} /G_{\rm M} \ll 1$ which means that the bulk conductance is much larger than the interface one, we finally obtain  
\begin{eqnarray}
  j_{\rm s}^{\rm M} (x) &=& 
  \frac{g_{\rm F/A} (\omega) V_{\rm s} e^{-(x-d_{\rm A})/\lambda_{\rm M}}}
  {1+ {G_{\rm F}}/{G_{\rm A/M}} }. 
  \label{eq:jsRES01}
\end{eqnarray}

In the next section, we use Eq.~(\ref{eq:jsRES01}) to discuss experimental results of Ref.~\cite{Qiu16}.

\begin{figure}[t] 
  \begin{center}
    \includegraphics[width=8.5cm]{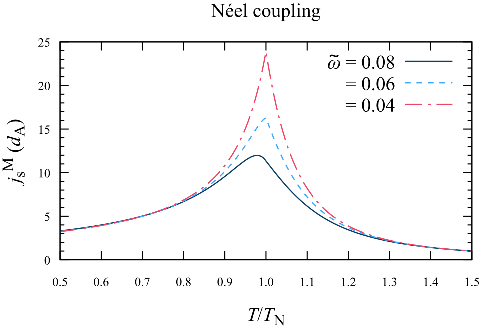} 
  \end{center}
  \caption{The same as in Fig.~\ref{fig:jm-temp01} but for the N\'{e}el coupling case. 
  }
  \label{fig:jn-temp01}   
\end{figure}

\section{Application to experiments \label{Sec:VI}}  

\begin{figure*} [t]
  \begin{center}
    \includegraphics[width=18cm]{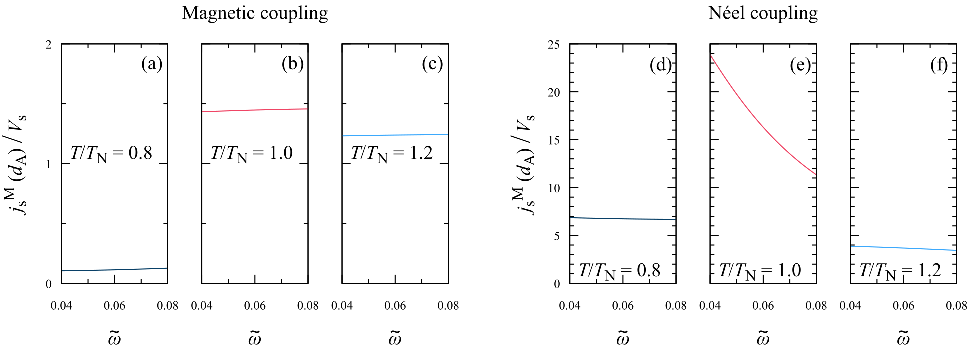} 
  \end{center}
  \caption{
    Angular frequency dependence of the pumped spin current density divided by the spin voltage, $j_{\rm s}^{\rm M}(d_{\rm A})/V_{\rm s}$, calculated from Eq.~(\ref{eq:jsRES01}) at several different temperatures. The magnetic coupling case [(a), (b), (c)] and the N\'{e}el coupling case [(d), (e), (f)] are shown. The data is normalized by its value at $T=1.5T_{\rm N}$, and $\widetilde{\Gamma}_m= 0.1$, $\widetilde{\Gamma}_n= 0.5$, and $K_0= 0.1$ are used. 
  }
  \label{fig:jn-omega01}   
\end{figure*}

In this section, we perform a numerical calculation for the theory and compare the results with experiments. Specifically, we focus on the spin pumping experiment in FI/AFI/M system~\cite{Qiu16}, where the spin pumping in a Y$_3$Fe$_5$O$_{12}$/CoO/Pt trilayer is investigated. In Ref.~\cite{Qiu16}, $T$-dependence of the spin pumping signal is observed to shows a peak at the N\'{e}el temperature $T_{\rm N}$ of CoO. Moreover, it is reported that the enhanced spin pumping signal peaked at $T_{\rm N}$ is strongly dependent on the external microwave frequency (see Fig.~4 therein). Below, we analyze these behaviors using the GL theory of the spin pumping developed in the previous sections. We argue that the strongly frequency-dependent enhancement of the spin pumping signal is a result of sizable N\'{e}el coupling at the Y$_3$Fe$_5$O$_{12}$/CoO interface. 

In order to examine the spin pumping in a FI/AFI/M trilayer, we use Eq.~(\ref{eq:jsRES01}) and numerically calculate the spin current density $j_{\rm s}^{\rm M}(d_{\rm A})$ pumped from FI into M through the AFI layer. In our numerical calculation, we measure magnetic field and angular frequency in units of $\mathfrak{h}_0$ and $\gamma \mathfrak{h}_0$, and define $\widetilde{H}_0= H_0/\mathfrak{h}_0$, $\widetilde{\omega}= \omega/\gamma \mathfrak{h}_0$, $\widetilde{\Gamma}_m= \Gamma_m/\gamma \mathfrak{h}_0$, and $\widetilde{\Gamma}_n= \Gamma_n/\gamma \mathfrak{h}_0$. Throughout our numerical calculations, we set $A= 1.0$ and $\Theta= 1.4$ for parameter $r_0$, and $D'= 10.0$. Note that in the following discussion $\omega$ satisfies the resonance condition 
\begin{equation}
  \omega= \gamma H_0 .
\end{equation}

Let us first discuss the case when the interfacical exchange interaction is given by the magnetic coupling [Eq.~(\ref{eq:magn-int01})]. Figure~\ref{fig:jm-temp01} shows the spin current density pumped into the M layer, $j_{\rm s}^{\rm M}(d_{\rm A})$, calculated from Eq.~(\ref{eq:jsRES01}) as a function of temperature. Because $j_{\rm s}^{\rm M}(d_{\rm A})$ depends on the angular frequency of ac magnetic field [Eq.~(\ref{eq:h_ac01})] through $g_{\rm F/A} (\omega)$, we plot $j_{\rm s}^{\rm M}(d_{\rm A})$ for several different choices of $\omega$. We see that the pumped current has a peak structure at $T=T_{\rm N}$~\cite{Okamoto16,Chen16}, but this peak looks more like a moderate cusp. Note that at first glance this cusp seems consistent with the experimental finding~\cite{Qiu16}, but in contrast to the experiment, the calculated height of the cusp at $T=T_{\rm N}$ remains unchanged while varying the values of angular frequency $\omega$.

Let us next discuss the case when the interfacical exchange interaction is given by the N\'{e}el coupling [Eq.~(\ref{eq:neel-int01})]. Figure~\ref{fig:jn-temp01} shows temperature dependence of $j_{\rm s}^{\rm M}(d_{\rm A})$, calculated from Eq.~(\ref{eq:jsRES01}) for several different values of ${\omega}$. We again see that the pumped current has a peak at $T=T_{\rm N}$. However, in this N\'{e}el coupling case, we find that the peak height at $T=T_{\rm N}$ has a strong angular frequency dependence as experimentally found in Ref.~\cite{Qiu16} (see Fig.~4 therein). 
We emphasize that, while a similar result to Fig.~\ref{fig:jm-temp01} that there appears a moderate cusp structure at $T=T_{\rm N}$ was found by previous theoretical approaches~\cite{Okamoto16,Chen16}, the result shown in Fig.~\ref{fig:jn-temp01} that the spin pumping signal has a strong angular frequency dependence near $T_{\rm N}$ has never been obtained before. In the next section, we explain the origin of the strong angular frequency dependence in terms of N\'{e}el order parameter fluctuations that manifest themselves through the interfacial N\'{e}el coupling.

To make the strong angular frequency dependence more visible, in Fig.~\ref{fig:jn-omega01}, we plot $j_{\rm s}^{\rm M}(d_{\rm A})$ for the magnetic coupling case [Figs.~\ref{fig:jn-omega01}(a)--\ref{fig:jn-omega01}(c)] and N\'{e}el coupling case [Figs.~\ref{fig:jn-omega01}(d)--\ref{fig:jn-omega01}(f)] as a function of angular frequency $\widetilde{\omega}$ at several different temperatures. In the magnetic coupling case [Figs.~\ref{fig:jn-omega01}(a)--\ref{fig:jn-omega01}(c)], we do not observe strong dependence on angular frequency even just at the N\'{e}el temperature $T=T_{\rm N}$ [Figs~\ref{fig:jn-omega01}(b)]. By contract, in the N\'{e}el coupling case [Figs.~\ref{fig:jn-omega01}(d)--\ref{fig:jn-omega01}(f)], although $\widetilde{\omega}$ dependence is weak at a temperature slightly shifted from the N\'{e}el temperature $T_{\rm N}$ [Figs.~\ref{fig:jn-omega01}(d) and \ref{fig:jn-omega01}(f)], we find a very strong $\widetilde{\omega}$ dependence exactly at $T_{\rm N}$ [Fig.~\ref{fig:jn-omega01}(e)]. Therefore, this strong angular frequency dependence appearing only in the vicinity of $T_{\rm N}$ explains the experimental result reported in Ref.~\cite{Qiu16} (see Fig.~4 therein). 

Before summarizing this section, we comment on the effect of slight tilting of the antiferromagnetic easy axis from the in-plane direction. Although the spin alignment in a real sample may not be perfectly parallel to the in-plane direction of FI/AFI/M interface ($\parallel {\bf \hat{z}}$), we assumed that the magnetic easy axis of the AFI layer is parallel to ${\bf \hat{z}}$. In order to see the effect of a slight tilting of the AFI easy axis from the $z$ axis, we performed a numerical simulation of the spin pumping using Eqs.~(\ref{eq:LLG01}), (\ref{eq:TDGL_m01}), and (\ref{eq:TDGL_n01}), and numerically evaluate the spin conductance $g_{\rm F/A}(\omega)$ by tilting the magnetic easy axis of the AFI layer from the $z$ axis as ${\bf \hat{u}}= \cos \theta {\bf \hat{z}} + \sin \theta {\bf \hat{x}}$. Then, for $\theta =5 \degree$ and $10 \degree$, we find only a negligibly small change in the signal, and the results shown in Figs.~\ref{fig:jm-temp01} and \ref{fig:jn-temp01} remain almost unchanged.

  We also comment on the influence of simultaneous presence of magnetic coupling $J_m$ and N\'{e}el coupling $J_n$, since it is natural to expect that we encounter such a situation in real experiments. Microscopically, $J_m$ ($J_n$) is proportional to the sum (difference) of the interfacial exchange couplings for the two antiferromagnetic sublattices~\cite{Tang24}. Therefore, when the interface possesses the symmetry under the sublattice interchange (e.g., compensated interface), $J_n$ vanishes. By contrast, when the interface breaks the symmetry under the sublattice interchange (e.g., uncompensated interface), we obtain nonzero $J_n$ as well as nonzero $J_m$. So far we discussed the effect of $J_m$ and $J_n$ separately, but additionally we investigate the effect of simultaneous presence of $J_m$ and $J_n$ using the numerical simulation described in the previous paragraph. Then, by numerically calculating the pumped spin current $j_{\rm s}^{\rm M}(d_{\rm A})$, we find that even for a rather small value of $|J_n/J_m| \!\agt\! 0.1$ the effect of the N\'{e}el coupling is visible. This is inferred from the large difference in the the size of the vertical axis between Figs.~\ref{fig:jm-temp01} and \ref{fig:jn-temp01}.

To summarize this section, we have examined temperature and frequency dependence of the spin pumping signal in a FI/AFI/M trilayer. When the interfacial exchange interaction is given by the magnetic coupling, we find a moderate cusp at the N\'{e}el temperature $T_{\rm N}$ with no visible frequency dependence, while when the interfacial exchange interaction is given by the N\'{e}el coupling, we find a pronounced peak at $T_{\rm N}$ with a strong frequency dependence. Comparing the present results with Fig.~4 of Ref.~\cite{Qiu16}, we conclude that the Y$_3$Fe$_5$O$_{12}$/CoO interface of Ref.~\cite{Qiu16} is dominated by the N\'{e}el coupling, giving rise to the strongly frequency-dependent enhancement of the spin pumping as shown in Fig.~\ref{fig:jn-omega01}.

\section{Discussion and Conclusion  \label{Sec:VII} }

In the previous section we have seen that, in the case of the interfacial magnetic coupling, the spin pumping in a FI/AFI/M trilayer has a moderate cusp at the N\'{e}el temperature $T_{\rm N}$ with no visible frequency dependence. By contrast, in the case of the interfacial N\'{e}el coupling, the spin pumping signal has a pronounced peak at $T_{\rm N}$ with a strong frequency dependence. The underlying physics behind this difference is explained as follows. 

First of all, it is important to notice that temperature and frequency dependence of the spin pumping in the FI/AFI/M trilayer calculated in the previous section are mainly determined by the spin conductance $g_{\rm F/A}(\omega)$ at the FI/AFI [Eqs.~(\ref{eq:gFA01}) and (\ref{eq:gFA02})],
\begin{empheq}[left={g_{\rm F/A}(\omega) \propto \empheqlbrace}]{alignat=2}
  & \frac{1}{\omega} \operatorname{Im} {\chi}_{\rm A} (\omega)
  &\;\;\; \text{for} \;\;\; & \text{magnetic coupling, } \nonumber \\
  & \frac{1}{\omega} \operatorname{Im} {\psi}_{\rm A} (\omega)
  &\;\;\; \text{for} \;\;\; & \text{N\'{e}el~coupling.}
  \nonumber 
\end{empheq}

Next, we recall that the right-hand side of the above equation is related to the fluctuations of $\bmm$ and $\bmn$. Indeed, from the classical limit of fluctuation-dissipation theorem~\cite{Chaikin-text}, we have the relation 
\begin{eqnarray}
  \frac{1}{\omega} \operatorname{Im} {\chi}_{\rm A} (\omega)
  &=& \frac{1}{4 \kB T } \bra \bra \delta m_\omega^- \delta m_{-\omega}^+ \ket \ket, \\
  \frac{1}{\omega} \operatorname{Im} {\psi}_{\rm A} (\omega)
  &=& \frac{1}{4 \kB T } \bra \bra \delta n_\omega^- \delta n_{-\omega}^+ \ket \ket,  
\end{eqnarray}
where $\bra \bra \cdots \ket \ket$ is defined by $\bra \delta m_\omega^- \delta m_{\omega'}^+  \ket = 2 \pi \delta(\omega+\omega') \bra \bra \delta m_\omega^- \delta m_{-\omega}^+ \ket \ket$. The above equations mean that, in the case of the magnetic coupling the spin pumping in the FI/AFI/M trilayer is affected by the magnetic fluctuations $\bra \bra \delta m_\omega^- \delta m_{-\omega}^+ \ket \ket$, whereas in the case of the N\'{e}el coupling it is dominated by the N\'{e}el fluctuations $\bra \bra \delta n_\omega^- \delta n_{-\omega}^+ \ket \ket$.

Then, in the vicinity of $T_{\rm N}$, there appears anomalies in the N\'{e}el fluctuations $\bra \bra \delta n_\omega^- \delta n_{-\omega}^+ \ket \ket$, since the N\'{e}el vector is the order parameter of the phase transition. Therefore, in the case of the interfacial N\'{e}el coupling, the spin pumping signal shows a pronounced peak at $T_{\rm N}$ with a strong frequency dependence. By contrast, there are no anomalies in the magnetic fluctuations $\bra \bra \delta m_\omega^- \delta m_{-\omega}^+ \ket \ket$, because the magnetization is not the order parameter of the phase transition. Therefore, in the case of the interfacial magnetic coupling, the spin pumping signal exhibits a moderate cusp at $T_{\rm N}$ with no visible frequency dependence. 

The above intuitive argument is confirmed more quantitatively by considering the spin pumping signal above $T_{\rm N}$, where there is no N\'{e}el order (i.e., $n_{\rm eq}=0$) but there is strong N\'{e}el correlations. In the case of the interfacial magnetic coupling, by setting $n_{\rm eq}=0$ in Eq.~(\ref{eq:ImCHI01}), we obtain 
\begin{eqnarray}
  g_{\rm F/A}(\omega) &\propto& 
  \frac{\Gamma_m }{(\omega- \gamma H_0)^2 
    + \big[ \Gamma_m  / \epsilon_0 v_0 {\chi}_{\rm A}(0) \big]^2}
  \label{eq:gFA03}  
\end{eqnarray}
for $T > T_{\rm N}$. From this equation we see that, at the magnetic resonance condition $\omega= \gamma H_0$, the spin pumping signal shows no dependence on $H_0$, and its temperature dependence comes from the static magnetic susceptibility ${\chi}_{\rm A} (0)$ that exhibits merely a weak cusp structure at $T_{\rm N}$. By contrast, in the case of the interfacial N\'{e}el coupling, by setting $n_{\rm eq}=0$ in Eq.~(\ref{eq:ImPSI01}), we have 
\begin{eqnarray}
  g_{\rm F/A}(\omega) &\propto& 
  \frac{\Gamma_n }{\omega^2
    + (\Gamma_n K)^2}
  \label{eq:gFA04}
\end{eqnarray}
for $T > T_{\rm N}$, where $K= K_0+ (T-T_{\rm N})/T_{\rm N}$, and we used $\epsilon_0 v_0 {\chi}_{\rm A}(0) K \ll 1$. If $K$ was not a small parameter ($K \approx 1$), then the above equation would not show strong dependence on magnetic field and temperature under the resonance condition $\omega= \gamma H_0$ because we are in an overdamped region $\omega/\Gamma_n \ll 1$ in this situation. However, when $K_0 \approx 0.1$, we are in a situation $\omega \sim \Gamma_n K $ near $T_{\rm N}$, such that the spin pumping signal exhibits a strong $H_0$ and $T$ dependence according to Eq.~(\ref{eq:gFA04}). 

To summarize this paper, we have developed GL theory of the spin pumping through an antiferromagnetic layer near the N\'{e}el temperature. By analyzing temperature and frequency dependence of the spin pumping signal in a FI/AFI/M trilayer, we have shown that, when the interaction at the FI/AFI interface is given by the magnetic coupling, the spin pumping signal has a moderate cusp at the N\'{e}el temperature $T_{\rm N}$ with no visible frequency dependence. By contrast, when the interaction at the FI/AFI interface is given by the N\'{e}el coupling, the spin pumping signal shows a pronounced peak at $T_{\rm N}$ with a strong frequency dependence. We have argued that the strongly frequency-dependent spin pumping signal observed in Ref.~\cite{Qiu16} is a manifestation of the N\'{e}el order parameter fluctuations that develop in the vicinity of the antiferromagnetic phase transition. Since the N\'{e}el coupling at the FI/AFI interface is expected to emerge from magnetically uncompensated interfaces~\cite{Daniels15,Tang24}, the present work has clarified the importance of interface engineering in developing antiferromagnetic spintronics.

\acknowledgments 
This work was financially suported by JSPS KAKENHI Grants No. JP23K23209 and No. JP23K21077.

\appendix

\section{Derivation of Eq.~(\ref{eq:delta_js01}) \label{Sec:App01}}

Here, we consider the spin current across the AFI/M layer that is driven by a non-equilibrium spin accumulation $\delta \sigma^z$ in the M layer. Below, this spin current density is denoted as $\delta j_{\rm s}$. 

We begin with the case when the exchange interaction at the AFI/M interface is given by the magnetic coupling $J'_m$ [Eq.~(\ref{eq:magn-int02})]. Since, the thermal agitation is necessary to discuss the spin conductance~\cite{Chen16} in the situation under consideration, the spin current is calculated from stochastic generalization of Eq.~(\ref{eq:js02}): 
\begin{equation}
  \delta j_{\rm s} = \frac{J'_{m}}{{\cal S}_{\rm A/M} \hbar} \int_{-\infty}^\infty \frac{d \omega}{2 \pi}
  \operatorname{Im} \bra \bra \delta m^-_\omega \delta \sigma^+_{-\omega} \ket \ket. 
  \label{eq:delta_js02}
\end{equation}
Linearizing the TDGL equations~(\ref{eq:TDGL_m01}) and (\ref{eq:TDGL_n01}) with respect to $\delta \bmm_\omega$ and $\delta \bmn_\omega$, projecting $\delta \bmm_\omega$ and $\delta \bmn_\omega$ onto $\delta m_\omega^-$ and $\delta n_\omega^-$, we obtain 
\begin{eqnarray}
(\omega-\widehat{\mathcal{A}}) 
\begin{pmatrix}
\delta m^{-}_\omega \\
\delta n^{-}_\omega \\
\end{pmatrix}
&=&
-\frac{J'_m}{\hbar}
\begin{pmatrix}
m_{\rm eq} - i \frac{\Gamma_m}{\gamma \mathfrak{h}_0}\\
n_{\rm eq} \\
\end{pmatrix} \delta \sigma^-_\omega
+
\begin{pmatrix}
i \xi^-_{\omega} \\
i \eta^{-}_{\omega} \\
\end{pmatrix}  \nonumber \\
&&
+\frac{J'_m}{\hbar}
\begin{pmatrix}
\delta m^{-}_{\omega} \\
\delta n^{-}_{\omega} \\
\end{pmatrix} \delta \sigma^z 
, 
\label{eq:append_TDGL_matrixAF01}
\end{eqnarray}
where $\delta \sigma^z$ in the above equation is not regarded as a fluctuating variable but as an external disturbance. Then, by operating $\widehat{\cal G}=(\omega- \widehat{\cal A})^{-1} $, we obtain $\delta m^-_\omega$ and $\delta n^-_\omega$. Likewise, for the dynamics of $\delta \sigma^+_\omega$, we have 
\begin{eqnarray}
  \left( \omega+ \frac{i}{\tau_{\rm M}} \right) \delta \sigma^+_\omega
  &=&
  \frac{J'_m}{\hbar}
  \left( i \frac{\chi_{\rm M}(0)}{\tau_{\rm M}/\hbar} + \delta \sigma^z \right)
  \delta m^+_\omega + i \zeta^+_\omega, \nonumber \\
  \label{eq:append_Bloch01}
\end{eqnarray}
where we assumed $\gamma H_0 \tau_{\rm M} \ll 1$. Note again that $\delta \sigma^z$ is regarded as an external disturbance. Then, substituting $\delta m^-_\omega$ and $\delta \sigma^+_{-\omega}$ into Eq.~(\ref{eq:delta_js02}) and performing the thermal average, we obtain
\begin{equation}
  \delta j_{\rm s} = - \frac{4 J'^2_m \kB T}{{\cal S}_{\rm A/M} \hbar^2 \epsilon_0 v_0 \tau_{\rm M} }
  {\cal L}_1 \, \delta \sigma^z, 
\end{equation}
where ${\cal L}_1$ is defined by
\begin{equation}
  \mathcal{L}_1 =
  \int_{-\infty}^\infty \frac{d \omega}{2 \pi}
  \Big( \Gamma_m |\mathcal{G}_1 (\omega)|^2 + \Gamma_n|\mathcal{G}_2 (\omega)|^2 \Big) 
    |g(\omega)|^2, 
\end{equation}
and $g(\omega) = (\omega+ i/\tau_{\rm M})^{-1} $. The remaining calculation is basically the same as that presented in Ref.~\cite{Yamamoto19}. Introducing $\sqrt{(a-d)^2+4bc}= \sqrt{Z}= X + i Y$ and $\Gamma_\pm = - \operatorname{Im}(d \pm a)$, $\mathcal{L}_1$ is calculated as
\begin{equation}
  \mathcal{L}_1 = \frac{\mathcal{N}_1 \tau^2_{\rm M}}{\mathcal{D} \sqrt{Z^*}}, 
\end{equation}
where
\begin{eqnarray}
  \mathcal{N}_1 &=&
  \sqrt{Z^*} \Big[ \Gamma_m (a^* -d )\big(XY+ \Gamma_+ (a^* - d^*) \big) \nonumber \\
  &&   + 2 \Gamma_+ b (\Gamma_m c + \Gamma_n b ) 
    \Big]. 
\end{eqnarray}
Then, expanding the above equation up to the second order with respect to $\Gamma_\pm$, we obtain
\begin{equation}
  \delta j_{\rm s} = -2 J'^2_m \frac{\kB T {\chi}_{\rm A}(0) \tau_{\rm M}}
         {\mathcal{S}_{\rm A/M} \hbar^2 } \delta \sigma^z .
         \label{eq:delta_js03}
\end{equation}

Next, we consider the case when the exchange interaction at the AFI/M interface is given by the N\'{e}el coupling $J'_n$ [Eq.~(\ref{eq:neel-int02})]. In this case, Eq.~(\ref{eq:delta_js02}) is replaced by
\begin{equation}
  \delta j_{\rm s} = \frac{J'_n}{{\cal S}_{\rm A/M} \hbar} \int_{-\infty}^\infty \frac{d \omega}{2 \pi}
  \operatorname{Im} \bra \bra \delta n^-_\omega \delta \sigma^+_{-\omega} \ket \ket,
    \label{eq:delta_js04} 
\end{equation}
and similarly Eq.~(\ref{eq:append_TDGL_matrixAF01}) is replaced by 
\begin{eqnarray}
(\omega-\widehat{\mathcal{A}}) 
\begin{pmatrix}
\delta m^{-}_\omega \\
\delta n^{-}_\omega \\
\end{pmatrix}
&=&
-\frac{J'_n}{\hbar}
\begin{pmatrix}
n_{\rm eq} \\
m_{\rm eq} - i \frac{\Gamma_n}{\gamma \mathfrak{h}_0}\\
\end{pmatrix} \delta \sigma_{\omega}^-
+
\begin{pmatrix}
i \xi^-_{\omega} \\
i \eta^{-}_{\omega} \\
\end{pmatrix}  \nonumber \\
&&
+\frac{J'_n}{\hbar}
\begin{pmatrix}
\delta n^{-}_{\omega} \\
\delta m^{-}_{\omega} \\
\end{pmatrix} \delta \sigma^z
, 
\label{eq:append_TDGL_matrixAF02}
\end{eqnarray}
and Eq.~(\ref{eq:append_Bloch01}) by 
\begin{eqnarray}
  \left( \omega+ \frac{i}{\tau_{\rm M}} \right) \delta \sigma^+_\omega
  &=&
  \frac{J'_n}{\hbar}
  \left( i \frac{\chi_{\rm M}(0)}{\tau_{\rm M}/\hbar} + \delta \sigma^z \right)
  \delta n^+_\omega + i \zeta^+_\omega. \nonumber \\
  \label{eq:append_Bloch02}
\end{eqnarray}
Then, proceeding in the same way as the magnetic coupling case, we obtain
\begin{equation}
  \delta j_{\rm s} = - \frac{4 J'^2_n \kB T}{{\cal S}_{\rm A/M} \hbar^2 \epsilon_0 v_0 \tau_{\rm M} }
  \mathcal{L}_2 \, \delta \sigma^z, 
\end{equation}
where $\mathcal{L}_2$ is defined by
\begin{equation}
  \mathcal{L}_2 =
  \int_{-\infty}^\infty \frac{d \omega}{2 \pi}
  \Big( \Gamma_n|\mathcal{G}_4 (\omega)|^2 + \Gamma_m |\mathcal{G}_3 (\omega)|^2 \Big) 
    |g(\omega)|^2. 
\end{equation}
After some algebra, $\mathcal{L}_2$ is transformed as 
\begin{equation}
  \mathcal{L}_2 = \frac{\mathcal{N}_2 \tau_{\rm M}^2}{\mathcal{D} \sqrt{Z^*}}, 
\end{equation}
where $\mathcal{N}_2$ is defined by 
\begin{eqnarray}
  \mathcal{N}_2 &=&
  \sqrt{Z^*} \Big[ \Gamma_n (d^* -a ) \big(XY+ \Gamma_+ (d^* - a^*) \big) \nonumber \\
  &&   + 2 \Gamma_+ c (\Gamma_n b + \Gamma_m c ) 
    \Big]. 
\end{eqnarray}
Then, expanding the above equation up to the second order with respect to $\Gamma_\pm$, we obtain
\begin{equation}
  \delta j_{\rm s} = -2 J'^2_n \frac{\kB T {\psi}_{\rm A}(0) \tau_{\rm M}}
         {\mathcal{S}_{\rm A/M} \hbar^2 } \delta \sigma^z .
         \label{eq:delta_js05}
\end{equation}

Now we decompose $\delta j_{\rm s}$ [Eqs.~(\ref{eq:delta_js03}) and (\ref{eq:delta_js05})] as 
\begin{equation}
  \delta j_{\rm s} = \delta j_{\rm s}^{\rm pump} - \delta j_{\rm s}^{\rm back}, 
\end{equation}
then we find $\delta j_{\rm s}^{\rm pump}=0$, and $\delta j_{\rm s}^{\rm back}$ can be summarized as given in Eq.~(\ref{eq:delta_js01}).




\end{document}